\documentclass[10pt,conference]{IEEEtran}

\usepackage[T1]{fontenc}
\usepackage{cite}
\usepackage{booktabs}
\usepackage{array}
\usepackage{multirow}
\usepackage{microtype}
\usepackage{placeins}
\usepackage{graphicx}
\usepackage{amsmath}
\usepackage{xspace}
\usepackage{tikz}
\usetikzlibrary{arrows.meta,positioning,fit,shapes,calc}
\usepackage{enumitem}
\usepackage[hidelinks]{hyperref}
\usepackage{url}

\hypersetup{
  pdftitle={FirmPilot: Evidence-Guided Multi-Agent Environment Recovery for IoT Firmware Rehosting},
  pdfauthor={Yanbing Shen, Fan Zhang, and Haitao Xu},
  pdfsubject={IoT firmware rehosting with evidence-guided multi-agent environment recovery},
  pdfkeywords={firmware rehosting, IoT security, emulation, multi-agent system, retrieval-augmented generation}
}

\newcommand{\tool}{\mbox{\textsc{FirmPilot}}\xspace}
\newcommand{\baseline}{FirmAE\xspace}
\newcommand{\stepnum}[1]{\tikz[baseline=(n.base)]\node[draw,circle,inner sep=0.6pt](n){\scriptsize #1};}
\newcommand{\Description}[1]{}
\newcommand{\agentdesignfigure}[4]{%
  \begin{figure}[!hbp]
    \centering
    \includegraphics[width=\columnwidth,trim=4.1pt 4.1pt 4.1pt 4.1pt,clip]{#1}
    \caption{#2}
    \label{#4}
    \Description{#3}%
  \end{figure}
}
\newcommand{\agentdesignfiguretop}[4]{%
  \begin{figure}[!t]
    \centering
    \includegraphics[width=\columnwidth,trim=4.1pt 4.1pt 4.1pt 4.1pt,clip]{#1}
    \caption{#2}
    \label{#4}
    \Description{#3}%
  \end{figure}
}
\setlist[itemize]{topsep=2pt,itemsep=1pt,parsep=0pt,partopsep=0pt}
\AtBeginDocument{
  
  \setlength{\textfloatsep}{4pt}
  \setlength{\floatsep}{4pt}
  \setlength{\intextsep}{4pt}
  \setlength{\dbltextfloatsep}{4pt}
  \setlength{\dblfloatsep}{4pt}
  \setlength{\abovecaptionskip}{1pt}
  \setlength{\belowcaptionskip}{0pt}

}

\begin{document}

\title{FirmPilot: Evidence-Guided Multi-Agent Environment Recovery for IoT Firmware Rehosting}

\author{\IEEEauthorblockN{Yanbing Shen, Fan Zhang, and Haitao Xu}
\IEEEauthorblockA{Zhejiang University}}

\maketitle

\begin{abstract}
Firmware rehosting executes firmware images in emulated environments such as QEMU to enable scalable dynamic analysis of Internet of Things (IoT) devices. In practice, rehosting pipelines remain fragile across diverse real-world firmware images, as reaching an externally observable execution state depends on tightly coupled artifacts spanning boot scripts, persistent configuration (e.g., NVRAM-like key--value state), and network setup. Template-driven frameworks often fail to accommodate long-tail vendor conventions, while unconstrained use of large language models (LLMs) risks unsupported modifications and irreproducible executions.

We introduce \tool, an evidence-guided multi-agent framework for environment recovery in firmware rehosting. \tool reformulates rehosting as iterative environment reconstruction in which a search agent grounds decisions through similarity-based retrieval, a planner coordinates execution-accepted transitions, and specialized agents recover filesystem/init artifacts, persistent state, and network exposure. Through repeated execution and evidence-grounded artifact deltas, the system resolves cross-layer dependencies across boot, state, and networking that otherwise prevent firmware executions from reaching a stable, externally reachable state in emulation.

Evaluated on the large-scale, real-world LFwC firmware corpus, \tool improves web-service reachability over \baseline from 25.49\% to 52.39\% and network reachability from 39.30\% to 71.93\%. The resulting rehosts raise the average number of detected services per firmware from 0.86 to 1.62 and support downstream analysis workflows, including RouterSploit interaction and protocol-aware fuzzing over recovered service surfaces. The evaluation shows that evidence- and feedback-grounded agent coordination improves rehosting success, service recovery, and downstream utility in automated firmware rehosting.

\end{abstract}


\begin{table*}[!t]
  \centering
  \caption{Representative Firmware Emulation Systems for IoT Security Analysis}
  \label{tab:landscape}
  {\scriptsize
  \setlength{\tabcolsep}{2.8pt}
  \renewcommand{\arraystretch}{1.03}
  \begin{tabular}{p{0.30\textwidth}p{0.16\textwidth}p{0.13\textwidth}p{0.10\textwidth}p{0.11\textwidth}>{\centering\arraybackslash}p{0.06\textwidth}}
    \toprule
    System & Fidelity & Emulation & Analysis & Automation & Hardware \\
    \midrule
    Firmadyne~\cite{chen2016towards}, \baseline~\cite{kim2020firmae} & System and services & OS emulation & Dynamic & Pipeline & No \\
    FirmGuide~\cite{liu2021firmguide}, PANDaWAN~\cite{angelakopoulos2024pandawan} & System and services & OS emulation & Dynamic & Diagnosis & No \\
    Jetset~\cite{johnson2021jetset}, Greenhouse~\cite{tay2023greenhouse}, FIRMWELL~\cite{qin2026firmwell} & Service and dependencies & User-space & Dynamic & Dependency-aware & No \\
    Avatar2~\cite{muench2018avatar2} & System and peripherals & Hybrid & Dynamic & Assisted & Yes \\
    \mbox{HALucinator~\cite{clements2020halucinator}, P2IM~\cite{feng2020p2im}, Fuzzware~\cite{scharnowski2022fuzzware}} & Peripherals and MMIO & Model emulation & Fuzzing & Models & No \\
    \tool{} (this work) & System and services & OS emulation & Dynamic & Agentic & No \\
    \bottomrule
  \end{tabular}}
\end{table*}

\section{Introduction}
\label{sec:intro}

IoT security analysis often starts from firmware images rather than source code or live devices. Analysts need to know which services actually start, which persistent configuration state gates them, and whether security tools can interact with the resulting runtime. Whole-system firmware rehosting answers this need by unpacking a firmware image, reconstructing a bootable environment, and executing it in QEMU so that network-facing behavior can be probed at scale~\cite{bellard2005qemu,chen2016towards,kim2020firmae}.

The practical barrier is that rehosting remains brittle on heterogeneous firmware corpora~\cite{fasano2021sok,gustafson2019toward,angelakopoulos2024pandawan}. A failed run is rarely caused by one missing command. Boot scripts, vendor wrappers, NVRAM-like key--value state, interface naming, and service launch order interact with each other; a change that makes the kernel boot may still leave the web stack disabled or bound to an unreachable interface. The available feedback is also partial because serial logs, reachability probes, and service snapshots expose symptoms rather than a complete specification of the original device environment.

This setting is well suited to LLM-based agents, which can interpret logs, scripts, and vendor-specific conventions. However, treating an LLM as an unconstrained operator is a poor fit for rehosting because edits affect boot-critical artifacts, success depends on repeated execution, and uncontrolled generation can make results irreproducible. A useful agentic system must therefore make model influence bounded and auditable. Decisions should be grounded in observed or retrieved evidence, actions should pass through typed interfaces, and claims should be validated by fixed probes rather than by model explanations.

We present \tool, an evidence-guided multi-agent framework for firmware rehosting. \tool treats rehosting as \emph{evidence-indexed environment reconstruction}. The system indexes runtime observations and retrieved evidence, maps them to typed transitions over boot artifacts, persistent state, and network exposure, and accepts a transition only after re-execution and probing. A \textsc{Search} agent retrieves firmware-specific evidence, a \textsc{Plan} agent schedules environment-recovery actions, and \textsc{File}, \textsc{NVRAM}, and \textsc{Network} agents materialize structured deltas through bounded action interfaces. This design turns model reasoning into a constrained control plane for state and exposure recovery.

Beyond basic reachability, \tool targets \emph{security-workflow fidelity}, namely service-facing, protocol-consistent behavior that lets standard tools execute and leave auditable artifacts in a controlled offline environment. This shifts evaluation from root-page response to whether recovered firmware executions can sustain service discovery, protocol-aware fuzzing, and RouterSploit interactions.

We evaluate \tool on 10,033 images from the large-scale, real-world LFwC firmware corpus and on the 1,122-image public \baseline benchmark. On the LFwC corpus, \tool improves web-service reachability over \baseline from 25.49\% to 52.39\% and network reachability from 39.30\% to 71.93\%; on the public benchmark, where \baseline already performs strongly, \tool raises Web reachability from 79.4\% to 82.2\%. Full-corpus ablations attribute the gain to planning, state synthesis, network exposure recovery, retrieval, and filesystem/init recovery, while a Claude Code baseline reaches only 5.43\% web-service reachability on the same corpus. Across all 5,256 web-reachable LFwC rehosts, downstream analyses enumerate services, validate HTTP/HTTPS fingerprints, support manual validation of 193 RouterSploit findings, and enable protocol-aware fuzzing.

We make the following contributions.
\begin{itemize}[leftmargin=*]
  \item We formulate firmware rehosting as evidence-indexed environment reconstruction under partial observability, where progress depends on resolving coupled boot, state, and network dependencies through execution-accepted transitions.
  \item We design \tool, an evidence-guided multi-agent environment-recovery loop with typed action interfaces for filesystem/init recovery, NVRAM state synthesis, and network exposure recovery.
  \item We conduct a large-scale evaluation on 10,033 LFwC images and the public \baseline benchmark, combining full-corpus ablations, a general-purpose coding-agent baseline, and downstream workflow analyses to show that \tool improves rehosting success and produces 5,256 web-reachable executions that support service discovery, manually validated RouterSploit findings, and protocol-aware fuzzing.
\end{itemize}

\section{Background}
\label{sec:background}

\subsection{Firmware Emulation}
\label{sec:background:landscape}

Firmware emulation executes firmware code in a controlled environment so analysts can observe runtime behavior without physical devices. IoT-security systems range from whole-system QEMU rehosting to user-space service and dependency rehosting, hybrid execution, and peripheral/MMIO modeling; surveys show that these choices trade off scale, automation, and fidelity~\cite{fasano2021sok, gustafson2019toward}. Table~\ref{tab:landscape} positions representative systems.

\begin{figure*}[!t]
  \centering
  \includegraphics[width=0.96\textwidth,height=0.255\textheight,keepaspectratio]{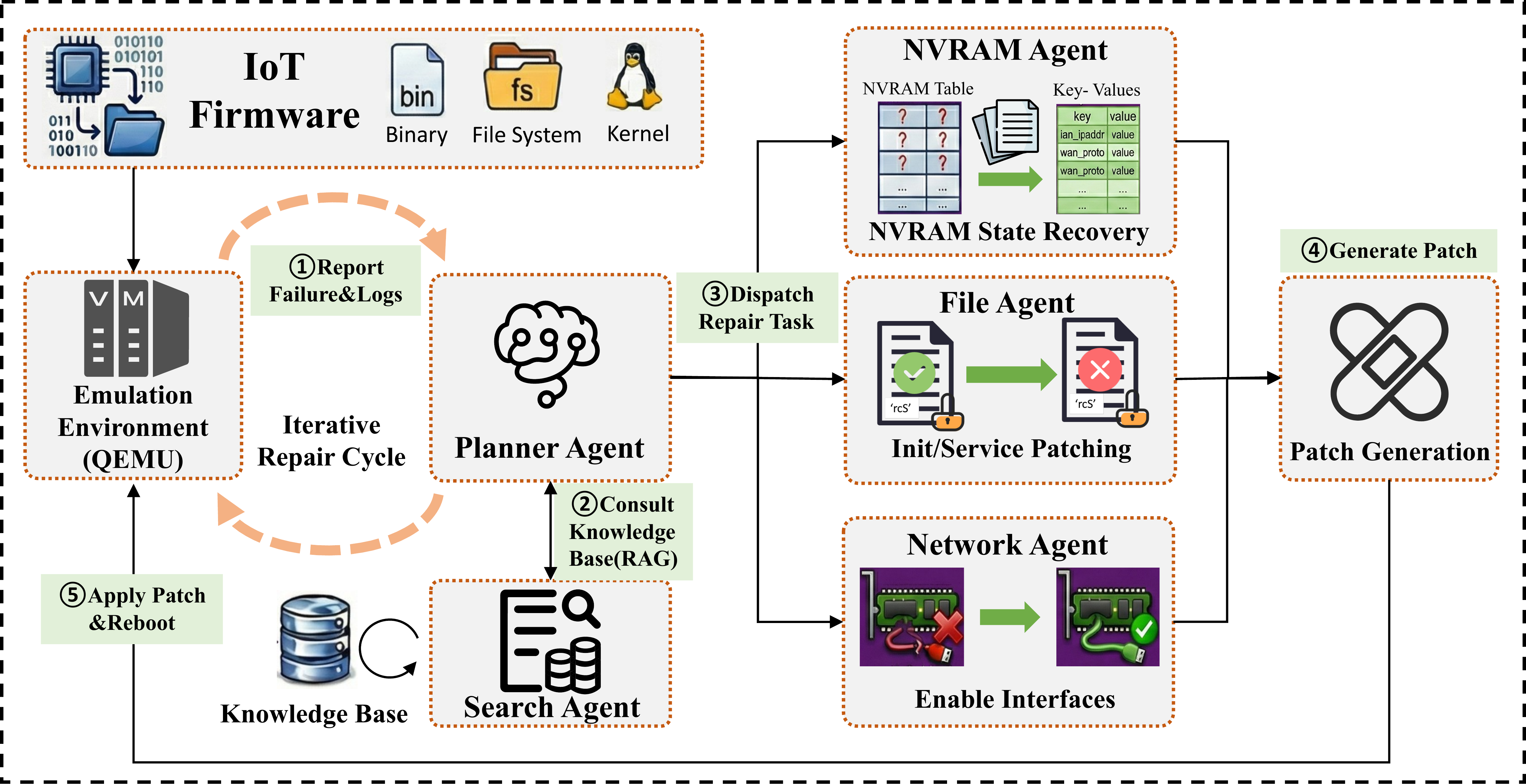}
  \caption{System Architecture of \tool}
  \Description{Architecture diagram of FirmPilot, showing how a Plan agent coordinates Search, File, NVRAM, and Network agents over a shared context and artifact store, with a QEMU backend providing logs and probe feedback in a closed loop.}
  \label{fig:architecture}
\end{figure*}

Whole-system pipelines remain the primary option for scalable service-facing analysis, but they are brittle on heterogeneous corpora because success depends on coupled init, persistent-state, and networking assumptions~\cite{fasano2021sok, gustafson2019toward}. Guidance systems improve observability, but they do not close the loop by converting evidence into bounded environment transitions~\cite{liu2021firmguide, angelakopoulos2024pandawan}. \tool therefore targets whole-system rehosting as evidence-indexed environment reconstruction.

\subsection{Rehosting Failure Modes}
\label{sec:background:rehosting}

Typical pipelines perform extraction, architecture detection, image construction, init selection, networking setup, and service probing. Each stage can fail when the emulator diverges from the device environment and runtime feedback is incomplete. Common failure patterns are as follows.
\begin{itemize}[leftmargin=*]
  \item \textbf{Boot and Init Discrepancies.} Incorrect init selection, divergent init systems, or vendor wrappers that assume device-specific mounts and environment.
  \item \textbf{Missing Persistent Configuration State.} Absent NVRAM-backed values for interface naming, addressing, service toggles, and authentication settings, preventing services from starting even when the kernel boots.
  \item \textbf{Network Configuration Discrepancies.} Inconsistent interface naming, bridging, or routes, leading to reachability without usable service exposure.
  \item \textbf{Cross-Binary Service Dependencies.} Multi-process service stacks that fail under incomplete state or partial startup, leading to fragile partial progress~\cite{redini2020karonte}.
\end{itemize}
These failures often cascade as a filesystem transition changes init progress, new logs reveal missing state keys, constrained defaults enable interface setup, and services then become observable under probing. Static one-shot pipelines do not exploit this feedback. User-space rehosting~\cite{tay2023greenhouse,johnson2021jetset,qin2026firmwell}, hybrid execution~\cite{zaddach2014avatar,muench2018avatar2}, and peripheral/MMIO modeling~\cite{clements2020halucinator,feng2020p2im} improve fidelity in complementary settings, but whole-system service analysis still needs scalable environment recovery.

\subsection{LLM Agent Foundations}
\label{sec:background:llm}

Large language models can interpret heterogeneous logs and scripts, but rehosting decisions must be grounded in verifiable signals and bounded in what they can change. We use retrieval-augmented prompting~\cite{lewis2020rag} and role-specific prompts for multi-step decisions~\cite{wei2022cot, shinn2024reflexion, li2023camel}. Unlike security agents for penetration-testing orchestration, fuzzing guidance, and program/binary analysis~\cite{deng2024pentestgpt, meng2024llmguided, li2023hitchhiker,liu2023llmbinary}, our agents act on boot-critical artifacts under strict budgets, so each role has a bounded action surface.

\section{Motivation and Design Challenges}
\label{sec:motivation}

Whole-system rehosting is the most direct path to scaling service-facing security workflows on heterogeneous IoT firmware, yet progress depends on iterative environment reconstruction across init, persistent state, networking, and services. A practical agentic system must interpret partial runtime evidence, apply small artifact deltas under fixed budgets, and avoid drift into unrealistic device state. This yields three design challenges.
\begin{itemize}[leftmargin=*]
  \item \textbf{C1--Limited Domain Knowledge in Firmware Emulation.} General-purpose models have uneven coverage of firmware rehosting toolchains and device-specific boot conventions, so plausible suggestions can still conflict with emulator constraints and lead to fragile fixes.
  \item \textbf{C2--Inference Under Partial and Noisy Observability.} Agents must infer the next environment transition from incomplete, vendor-specific logs, probes, and filesystem evidence while avoiding unsupported causal claims.
  \item \textbf{C3--Constrained and Convergent Environment Actuation.} Actions can mutate boot scripts, configuration files, and emulator arguments, so deltas must be typed, and the loop must converge under strict budgets without drifting toward unrealistic state injection.
\end{itemize}

These challenges motivate a separation between semantic inference and environment transition. In firmware rehosting, the unknown object is not a source-level patch but an execution environment whose state is only partially observable through logs and probes. \tool therefore treats LLM outputs as hypotheses over a constrained transition space, while state changes are performed by typed operators and accepted through execution evidence. This turns environment recovery into an evidence-conditioned search problem in which the model helps identify likely missing dependencies, the system exposes only the corresponding action surface, and progress is attributed to observed state changes, not model rationales.

\section{System Design}
\label{sec:method}

\subsection{Overview}
\label{sec:design:overview}

Figure~\ref{fig:architecture} illustrates \tool as an iterative environment-recovery loop over a shared context. Each iteration executes the current artifacts in QEMU, collects serial logs and probe results (\textbf{\stepnum{1}}), retrieves firmware-specific evidence (\textbf{\stepnum{2}}), plans a dependency-aware transition schedule (\textbf{\stepnum{3}}), and asks specialized agents to emit typed deltas (\textbf{\stepnum{4}}). The orchestrator applies accepted deltas, reboots the firmware, and reruns fixed probes (\textbf{\stepnum{5}}). The loop addresses \textbf{C1} through retrieval, \textbf{C2} through execution-grounded planning, and \textbf{C3} through budgeted, typed, and replayable actuation.

\subsection{Environment-Recovery Loop Formalization}
\label{sec:design:formal}

We model \tool as a bounded transition system. At iteration $t$, artifact state $x_t$ contains filesystem overlays, launch configuration, NVRAM overrides, and network exposure scripts. Executing $x_t$ in QEMU yields observations $o_t$ such as serial logs, reachability results, and service snapshots. The orchestrator extracts evidence $E_t$ and maintains $c_t=\langle x_{\le t}, o_{\le t}, E_{\le t}\rangle$; \textsc{Search} augments this context and \textsc{Plan} selects an ordered environment-recovery plan $p_t$.

Each action agent $a\in p_t$ models a bounded function from context to an artifact delta, $a(c_t)=\delta_{t,a}$. Applying these deltas yields the next artifact state.
\[
x_{t+1} \gets x_t \oplus \bigoplus_{a\in p_t} \delta_{t,a}.
\]
The operator $\oplus$ applies a delta to the current artifacts, and $\bigoplus_{a\in p_t}$ denotes sequential composition in plan order. The agent set is
\[
\mathcal{A}=\{\textsc{Search},\textsc{Plan},\textsc{File},\textsc{NVRAM},\textsc{Network}\},
\]
where \textsc{Search} and \textsc{Plan} acquire evidence and schedule actions, while \textsc{File}, \textsc{NVRAM}, and \textsc{Network} update $x_t$ through filesystem/init deltas, NVRAM overlays, and network artifacts. Model outputs are treated as candidate environment transitions rather than direct authority over the firmware image. A transition becomes part of the run state only after it is emitted through a typed interface, rendered by deterministic tooling, and validated by the next execution and fixed probes. The loop stops when Web reachability is achieved, no admissible action remains under unchanged evidence, or the outer budget is exhausted. The context records artifact versions, observations, retrieved support, and accepted deltas as an execution-indexed causal trace, preserving attribution between evidence, action, and outcome even when progress is non-monotonic.

\subsection{Search Agent for Evidence Acquisition}
\label{sec:design:search}

\agentdesignfigure{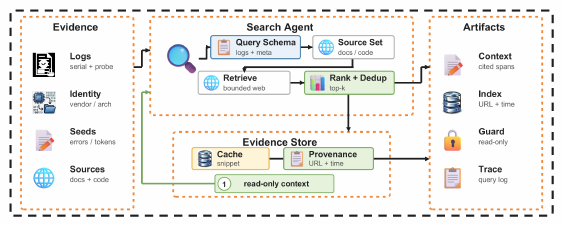}{Design of the \textsc{Search} Agent}{A system-design diagram showing the Search agent as a read-only evidence-acquisition component that abstracts runtime signals into structured queries, retrieves external references, records provenance, maintains a local evidence index, and emits cited context packets without mutating firmware artifacts.}{fig:search_design}

\textbf{Role and Evidence Flow.} The \textsc{Search} agent gives the loop a controlled path to outside knowledge. Its query packet is built from firmware metadata, vendor/model/board identifiers, architecture, recent serial-log windows, probe summaries, missing binary or library names, NVRAM key names, service binaries, port/banner observations, and compact failure tokens. Retrieval targets Web and GitHub-visible evidence such as vendor documentation, public code repositories, support/forum threads, and prior snippets cached by provenance. Retrieved resources enter the shared context only as cited snippets with query, URL, timestamp, source type, and similarity score. Search keeps a bounded candidate set and admits only the highest-ranked cited snippets to the shared context. It is read-only and cannot change files, state keys, or network parameters. This makes external evidence available to the planner without turning retrieval into an environment-transition mechanism~\cite{lewis2020rag}.

\subsection{Plan Agent for Orchestration and Scheduling}
\label{sec:design:plan}

\agentdesignfigure{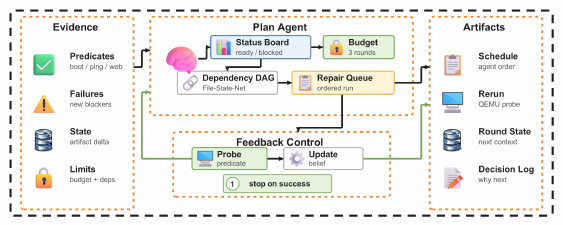}{Design of the \textsc{Plan} Agent}{A system-design diagram showing the Plan agent as a bounded orchestrator that converts runtime state and probe observations into progress beliefs and failure hypotheses, applies dependency and budget gates, dispatches recovery agents, and uses rerun feedback for the next round.}{fig:plan_design}

\textbf{Responsibilities and Feedback.} The \textsc{Plan} agent is the loop control point. It consumes artifact readiness, init and service candidates, probe outcomes, recent error signatures, retrieved snippets, port signals, and accepted deltas, then emits an ordered schedule over \textsc{File}, \textsc{NVRAM}, and \textsc{Network} with a failure class and stop/retry condition. The schedule is explicit because persistent-state recovery can enable service startup, and exposure recovery is meaningful only after a daemon binds or a likely bind address appears. Actions execute through typed interfaces. \textsc{File} emits entrypoint and startup deltas, \textsc{NVRAM} emits key/value/source/guard tuples, and \textsc{Network} emits IP, interface, bridge/VLAN, port, and QEMU-argument tuples. The planner never writes artifacts directly; the next plan observes updated logs and probe state after emulation-relevant mutations.

\begin{table*}[t]
  \centering
  \caption{Agent Roles and Bounded Action Interfaces in \tool}
  \label{tab:agent_contracts}
  {\scriptsize
  \setlength{\tabcolsep}{0pt}
  \renewcommand{\arraystretch}{1.16}
  \begin{tabular}{@{}>{\raggedright\arraybackslash}p{0.10\textwidth}@{\hspace{0.018\textwidth}}>{\raggedright\arraybackslash}p{0.27\textwidth}@{\hspace{0.018\textwidth}}>{\raggedright\arraybackslash}p{0.27\textwidth}@{\hspace{0.018\textwidth}}>{\raggedright\arraybackslash}p{0.27\textwidth}@{}}
    \toprule
    \textbf{Agent} & \textbf{Evidence used} & \textbf{Bounded action} & \textbf{Execution guard} \\
    \midrule
    \textsc{Search} & Serial/probe logs; firmware metadata; service/failure tokens & Bounded Web/GitHub queries; top-5 cited snippets & Provenance logging; retrieval cache only; no artifact mutation. \\
    \textsc{Plan} & Probe state; artifact readiness; recent errors; retrieved snippets & Agent sequence; failure class; stop/retry state & Budget and dependency checks; no direct filesystem, state, or network mutation. \\
    \textsc{File} & Root filesystem; init candidates; service binaries; retrieval context & Entrypoint; service command; filesystem/startup delta & Path checks; command filtering; replayable startup deltas. \\
    \textsc{NVRAM} & Access traces; configuration files; missing-key evidence & Key, value, source, and guard overlays & Value filtering; malformed entries removed; overlays isolated from rootfs edits. \\
    \textsc{Network} & Boot logs; listener/interface evidence; probe outcomes & IP, interface, bridge/VLAN, port, and QEMU-argument tuples & Consistency checks; rendering from accepted exposure deltas; fixed reachability probes. \\
    \bottomrule
  \end{tabular}
  }
\end{table*}

Table~\ref{tab:agent_contracts} defines the role contract. \textsc{Search} and \textsc{Plan} reason over evidence and ordering, while \textsc{File}, \textsc{NVRAM}, and \textsc{Network} enact typed transitions over boot artifacts, persistent state, and exposure state.

\subsection{\texorpdfstring{File Agent for Boot-Environment Recovery}{File Agent for Boot-Environment Recovery}}
\label{sec:design:file}

\textbf{Objective and Artifacts.} The \textsc{File} agent recovers boot-time filesystem artifacts so static firmware contents reach a runnable user space. It maintains candidate init entrypoints and service-launch specifications, emitting minimal filesystem deltas plus replayable records of inferred init paths, startup commands, and evaluated variants.

\textbf{Discovery and Service Selection.} Figure~\ref{fig:file_design} separates boot-environment inference from artifact mutation. The agent ranks init candidates from filesystem enumeration, common init locations, kernel command-line hints, and retrieved evidence. The LLM may add vendor-specific hypotheses, but it must express them as typed startup deltas and may abstain when evidence is insufficient. \tool validates paths, normalizes arguments, rejects shell metacharacters and destructive utilities, and defaults to existing on-image init scripts when external evidence is insufficient.

\agentdesignfigure{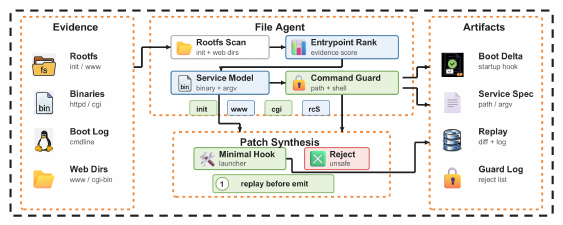}{Design of the \textsc{File} Agent}{A system-design diagram showing the File agent as a boot-environment reconstruction component that converts rootfs layout and boot evidence into an init/service model, validates paths and commands, emits boot deltas and service specifications, and closes the loop through service-startup feedback.}{fig:file_design}

\subsection{NVRAM Agent for Persistent State Synthesis}
\label{sec:design:nvram}

\agentdesignfigure{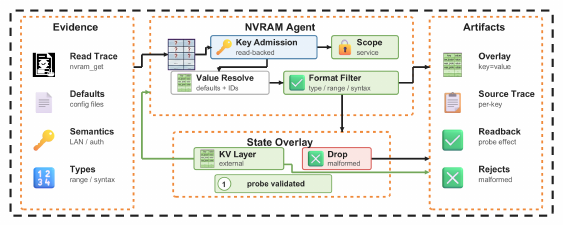}{Design of the \textsc{NVRAM} Agent}{A system-design diagram showing the NVRAM agent as a persistent-state synthesis component that derives key dependencies from read traces and configuration defaults, resolves or abstains on values, filters formats, isolates state from rootfs mutation, and validates the resulting overlay through service probes.}{fig:nvram_design}

\textbf{Purpose and Motivation.} The \textsc{NVRAM} agent reconstructs persistent configuration dependencies required for service startup. Embedded services consult NVRAM-like stores for interface naming, access control, feature flags, and management-service settings; missing keys can keep daemons disabled even after boot and network recovery. Reconstructed state is materialized as an external key--value overlay rather than a root-filesystem mutation.

\textbf{Grounded State Synthesis.} Figure~\ref{fig:nvram_design} shows the state-recovery boundary. The agent does not apply a universal NVRAM template; it treats persistent state as a per-firmware inference problem. It builds candidates from runtime NVRAM reads, default files, configuration strings, and model/board/serial/MAC/LAN/Web evidence, then asks the LLM to resolve only prioritized keys not already determined by firmware evidence. Each proposal must be expressed as a key, value, source, and guard. The action interface preserves identity values, rejects malformed fragments and broad credential-series artifacts, and drops empty entries before producing the runtime overlay. State recovery runs when execution evidence shows missing NVRAM devices, invalid flash configuration, pre-bind crashes, or missing product identity; the transition is accepted only when the next execution supports it.

\subsection{\texorpdfstring{Network Agent for Connectivity and Exposure Recovery}{Network Agent for Connectivity and Exposure Recovery}}
\label{sec:design:network}

\agentdesignfiguretop{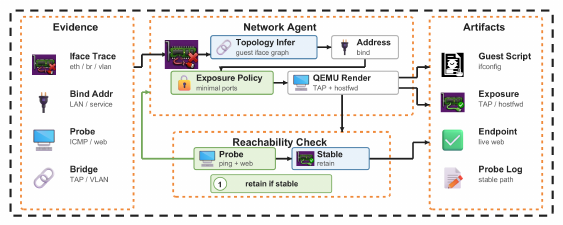}{Design of the \textsc{Network} Agent}{A system-design diagram showing the Network agent as an exposure-layer recovery component that derives interface topology from logs and probes, forms address and service-exposure hypotheses, applies consistency and minimal-change guards, renders guest scripts and QEMU arguments, and validates them through reachability probes.}{fig:network_design}

\textbf{Purpose and Design Goal.} The \textsc{Network} agent recovers the exposure layer needed for firmware-started services to become reachable, including guest-side connectivity and host-side forwarding/TAP state. It maps noisy logs and vendor-specific interface conventions into a structured network representation, then selects the smallest evidence-consistent update that improves reachability while preserving application behavior.

\textbf{Inference, Exposure, and Validation.} Figure~\ref{fig:network_design} separates exposure inference from environment mutation. The agent reasons over interfaces, addresses, VLANs, bridges, MAC transitions, and service bind addresses using serial logs, probes, NVRAM LAN evidence, and the \baseline-compatible seed. It then emits a typed exposure delta rather than a broad forwarding rule. The action interface materializes only the selected IP/interface, bridge/VLAN, port, and QEMU-argument tuples. DHCP-like guests use user-mode forwarding for observed bind ports, TAP mode preserves guest subnets and exported probe IPs, and ARM runs keep one active interface with additional probe IPs for multi-address stacks. Updates are accepted only when re-execution confirms both network reachability and a live service endpoint under the fixed probes.

\FloatBarrier

\section{Experimental Evaluation}
\label{sec:evaluation}

\begin{table*}[!t]
  \centering
  \caption{Vendor-Level Ping and Web-Service Reachability on the Public \baseline Benchmark and LFwC}
  \label{tab:rq1_web_by_vendor}
  {\scriptsize
  \setlength{\tabcolsep}{2.55pt}
  \renewcommand{\arraystretch}{1.03}
  \resizebox{0.98\textwidth}{!}{%
  \begin{tabular}{@{}lrrrrr|lrrrrr@{}}
    \toprule
    \multicolumn{6}{c|}{\textbf{Public \baseline benchmark}} &
    \multicolumn{6}{c}{\textbf{LFwC}} \\
    \cmidrule(lr){1-6}
    \cmidrule(l){7-12}
    \multirow{2}{*}{\raisebox{-0.45ex}{\textbf{Vendor}}} & \multirow{2}{*}{\raisebox{-0.45ex}{\textbf{Images}}} &
    \multicolumn{2}{c}{\textbf{Ping}} &
    \multicolumn{2}{c|}{\textbf{Web service}} &
    \multirow{2}{*}{\raisebox{-0.45ex}{\textbf{Vendor}}} & \multirow{2}{*}{\raisebox{-0.45ex}{\textbf{Images}}} &
    \multicolumn{2}{c}{\textbf{Ping}} &
    \multicolumn{2}{c}{\textbf{Web service}} \\
    \cmidrule(lr){3-4}
    \cmidrule(lr){5-6}
    \cmidrule(lr){9-10}
    \cmidrule(l){11-12}
    & & \baseline & \tool & \baseline & \tool &
    & & \baseline & \tool & \baseline & \tool \\
    \midrule
    Netgear & 375 & 351 (93.6\%) & 356 (94.9\%) & 336 (89.6\%) & 345 (92.0\%) &
    Netgear & 2,551 & 1,337 (52.41\%) & 2,254 (88.36\%) & 879 (34.46\%) & 1,679 (65.82\%) \\
    D-Link & 262 & 249 (95.0\%) & 249 (95.0\%) & 231 (88.2\%) & 236 (90.1\%) &
    D-Link & 1,909 & 175 (9.17\%) & 726 (38.03\%) & 130 (6.81\%) & 574 (30.07\%) \\
    TP-Link & 148 & 148 (100.0\%) & 148 (100.0\%) & 113 (76.4\%) & 119 (80.4\%) &
    ASUS & 1,645 & 871 (52.95\%) & 1,310 (79.64\%) & 757 (46.02\%) & 1,161 (70.58\%) \\
    Trendnet & 118 & 101 (85.6\%) & 101 (85.6\%) & 73 (61.9\%) & 78 (66.1\%) &
    Ubiquiti & 1,394 & 312 (22.38\%) & 897 (64.35\%) & 13 (0.93\%) & 396 (28.41\%) \\
    ASUS & 107 & 63 (58.9\%) & 64 (59.8\%) & 62 (57.9\%) & 64 (59.8\%) &
    TP-Link & 1,163 & 723 (62.17\%) & 1,129 (97.08\%) & 468 (40.24\%) & 817 (70.25\%) \\
    Linksys & 55 & 48 (87.3\%) & 49 (89.1\%) & 44 (80.0\%) & 48 (87.3\%) &
    Trendnet & 752 & 379 (50.40\%) & 565 (75.13\%) & 268 (35.64\%) & 381 (50.66\%) \\
    Belkin & 37 & 30 (81.1\%) & 31 (83.8\%) & 22 (59.5\%) & 22 (59.5\%) &
    AVM & 301 & 43 (14.29\%) & 83 (27.57\%) & 16 (5.32\%) & 39 (12.96\%) \\
    Zyxel & 20 & 18 (90.0\%) & 18 (90.0\%) & 10 (50.0\%) & 10 (50.0\%) &
    Linksys & 173 & 6 (3.47\%) & 110 (63.58\%) & 0 (0.00\%) & 92 (53.18\%) \\
    \multicolumn{6}{c|}{} &
    EnGenius & 145 & 97 (66.90\%) & 143 (98.62\%) & 26 (17.93\%) & 117 (80.69\%) \\
    \midrule
    \textbf{Overall} & \textbf{1,122} & \textbf{1,008 (89.8\%)} & \textbf{1,016 (90.6\%)} & \textbf{891 (79.4\%)} & \textbf{922 (82.2\%)} &
    \textbf{Overall} & \textbf{10,033} & \textbf{3,943 (39.30\%)} & \textbf{7,217 (71.93\%)} & \textbf{2,557 (25.49\%)} & \textbf{5,256 (52.39\%)} \\
    \bottomrule
  \end{tabular}%
  }
  }
\end{table*}

We evaluate \tool as an environment-aware firmware rehosting system. The experiments ask four questions about end-to-end effectiveness, mechanism contribution, domain specificity, and downstream workflow evidence.
\begin{itemize}[leftmargin=*]
  \item \textbf{RQ1 Rehosting effectiveness.} Does \tool improve network and web-service reachability over a widely used automated rehosting baseline on a large firmware corpus?
  \item \textbf{RQ2 Mechanisms.} Which environment-recovery mechanisms account for the observed improvement, and are their effects measurable at full-corpus scale?
  \item \textbf{RQ3 Domain specificity.} Can a general-purpose coding agent substitute for a firmware-specific recovery workflow?
  \item \textbf{RQ4 Downstream security workflow support.} Do web-reachable \tool rehosts support downstream security workflows beyond reachability?
\end{itemize}

\textbf{Dataset and baselines.} The primary large-scale dataset is LFwC~\cite{helmke2024lfwc}, a public Linux-firmware corpus built for reproducible firmware vulnerability research with documented acquisition metadata, unpacking checks, deduplication, content identification, and ground-truth annotations. Its metadata snapshot contains 10,913 records; 880 vendor or archival links were no longer reachable, leaving 10,033 locally executable images. These 10,033 images are the fixed denominator for all main LFwC comparisons, ablations, and general-agent baselines. We compare \tool against \baseline~\cite{kim2020firmae}, a QEMU-based rehosting pipeline from the Firmadyne lineage~\cite{chen2016towards} that produces comparable disk images, run scripts, logs, and probes.

We also evaluate on the public \baseline benchmark, which contains 1,122 images across 8 vendors and ARM-LE/MIPS-BE/MIPS-LE architectures. This established benchmark is closely aligned with \baseline's original templates and provides a comparability point, while LFwC supplies the main corpus.

\textbf{Comparison and ablation protocol.} The evaluation separates three sources of evidence, namely a deterministic rehosting baseline, an internal role ablation, and a general-purpose coding-agent baseline. \baseline represents the template-driven deterministic alternative, with fixed image construction, boot inference, NVRAM defaults, network heuristics, QEMU execution, and the same probes, but no retrieval, planning, or typed agent actions. RQ1 measures the deterministic-to-\tool gap under identical predicates. RQ2 keeps \tool's runnable substrate while replacing one adaptive role at a time with its deterministic fallback. In this setting, no-Search uses only local evidence, no-Plan follows a fixed dependency order, no-File uses baseline-style init and launch candidates, no-NVRAM retains FirmAE-compatible defaults, and no-Network retains deterministic network inference. RQ3 replaces the specialized action space with a general coding agent under the same dataset, timeout, input evidence, and success predicates.

\textbf{Execution policy and metrics.} All RQ1--RQ3 systems use a QEMU-based execution path~\cite{bellard2005qemu}, fixed probes, and identical success predicates. All agents use DeepSeek V4 Flash as the LLM backend throughout the evaluation. Consistent with FirmAE's Docker evaluation workflow, which uses a 2,400-second per-firmware emulation-result check, each RQ1--RQ3 image receives the same 2,400-second wall-clock budget for rehosting and probing; \baseline, \tool, and Claude Code all use this cap, and \tool runs at most three planner rounds within it. \emph{Network reachable} requires a standardized ICMP response. \emph{Web-service reachable} requires a completed HTTP(S) transaction to an exported guest-IP candidate, using bounded GET requests to \texttt{/}. HTTPS candidates complete the TLS handshake before request delivery, and the probe records status, redirects, authentication challenges, banners, and fingerprints when present. Success includes root pages, redirects, authentication gates, and other parseable management responses; status \texttt{200} is not required. Each request uses a two-second timeout inside the image budget.

The predicates separate network liveness from service readiness. ICMP measures packet exchange, while Web success requires a reachable and parseable management endpoint. This keeps Ping/Web columns directly comparable across \baseline, \tool, and the general-agent baseline.

RQ4 evaluates whether the recovered rehosts can support downstream analysis workflows beyond basic reachability. The audit reruns service discovery with \texttt{nmap -O -sV}, retries incomplete scans with \texttt{nmap -Pn -n -sV}, and credits valid scans only when complete XML contains auditable service records. HTTP/HTTPS evidence requires an endpoint, service label, protocol, and available banner/fingerprint fields from nmap or confirmation probes. RouterSploit and protocol-aware fuzzing then operate on this recovered surface. Counts use the 10,033-image denominator unless a table states otherwise.

\textbf{Evaluation records.} All reported counts are computed from per-image execution records rather than ad hoc log inspection. Each record binds the firmware identifier, execution budget, generated artifacts, serial output, probe transcript, and final reachability labels. RQ4 records additionally attach the service-discovery, RouterSploit, and fuzzing artifacts credited in the downstream workflow. This common record structure keeps the LFwC, ablation, and general-agent comparisons aligned at the image level.

\subsection{RQ1 Overall Rehosting Effectiveness}
\label{sec:evaluation:rq1}

\textbf{Overall reachability.} RQ1 tests the central rehosting claim that a rehost becomes useful for interactive analysis only when the firmware exposes an externally reachable service, not merely when the kernel boots or a network stack responds. Under identical probes on the full 10,033-image LFwC set, \baseline makes 3,943 images network reachable and 2,557 images web-service reachable, corresponding to 39.30\% and 25.49\% of the corpus. \tool increases these numbers to 7,217 network-reachable images and 5,256 web-service-reachable images, corresponding to 71.93\% and 52.39\%. This is an absolute improvement of 3,274 network-reachable images and 2,699 web-service-reachable images over \baseline; Table~\ref{tab:rq1_web_by_vendor} reports the public-benchmark and LFwC vendor results under the same Ping/Web-service columns.

\textbf{Benchmark and long-tail effects.} On the public \baseline benchmark, where existing templates already cover many cases, \tool preserves the high reachability level expected on this established setting and raises Web-service reachability from 79.4\% to 82.2\%. The LFwC results expose a larger gap under heterogeneous, long-tail firmware images. For Netgear, ASUS, TP-Link, Linksys, and EnGenius, \tool reaches web-service rates above 50\%. For D-Link, Ubiquiti, and AVM, the absolute rates remain lower, but \tool still substantially improves over \baseline, especially on Ubiquiti where \baseline reaches only 13 web services while \tool reaches 396. This contrast shows that environment recovery is most valuable when template coverage is sparse and reachability depends on coordinating boot, persistent state, and network exposure.

\begin{figure}[t]
  \centering
  \includegraphics[width=0.90\linewidth]{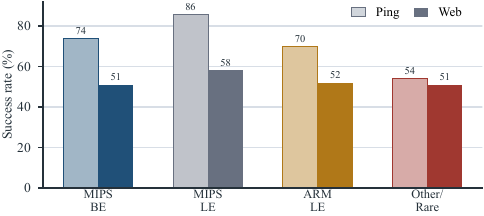}
  \caption{LFwC Ping/Web Success by Firmware Architecture}
  \label{fig:rq1_arch_success}
\end{figure}

\textbf{Architecture split.} Figure~\ref{fig:rq1_arch_success} complements the vendor view with an endian-aware primary firmware architecture split of the same 10,033 LFwC images. \tool reaches Web services on 1,588 MIPS-BE, 942 MIPS-LE, and 2,192 ARM-LE images, while making 2,309, 1,389, and 2,951 images network-reachable in the same groups. These router-class architecture groups account for 4,722 of the 5,256 Web-service successes; PowerPC, x86, unresolved-endian, and rare groups are aggregated into the long tail.

\begin{figure}[t]
  \centering
  \includegraphics[width=0.92\linewidth]{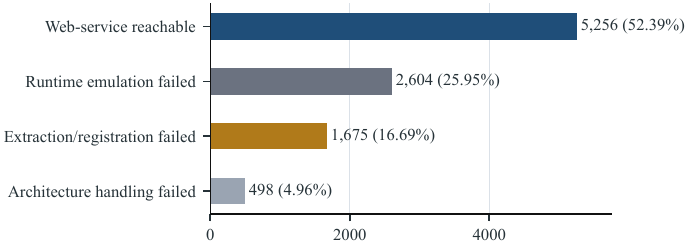}
  \caption{Final-state distribution for the full \tool run on LFwC.}
  \label{fig:rq1_outcomes}
\end{figure}

\textbf{Residual failure modes.} The final-state distribution in Figure~\ref{fig:rq1_outcomes} shows that \tool's failures are not dominated by one residual bottleneck. Among the 10,033 images, 5,256 reach a web service. The remaining images primarily fail because of runtime emulation failures (2,604 images), extraction or registration failures (1,675 images), and architecture-handling failures (498 images). This distribution shows that the remaining failures are spread across extraction, architecture handling, runtime execution, and late-stage service launch rather than concentrated in one probe or service check.

\textbf{Interpretation.} The RQ1 result is strongest on the large LFwC corpus because LFwC contains vendor and version diversity that stresses fixed templates. \tool does not merely increase Ping success and then inherit Web success automatically. The absolute Web gain of 2,699 images is close to the network gain in scale but not identical in source, since many additional successes require persistent-state synthesis for service toggles and exposure recovery for bind-address mismatches after basic boot progress. The vendor-level table makes this visible. Large gains on Ubiquiti, Linksys, EnGenius, and D-Link indicate that environment reconstruction exposes service paths left dormant by static baselines.

\subsection{RQ2 Contribution of Specialized Agents}
\label{sec:evaluation:rq2}

\textbf{Ablation setup.} RQ2 evaluates specialized roles after holding the deterministic execution substrate fixed. The deterministic alternative is template-driven rehosting represented by \baseline, which fixes image construction, boot inference, NVRAM defaults, and network setup through heuristic templates. We run leave-one-module ablations for SearchAgent, PlanAgent, FileAgent, NVRAMAgent, and NetworkAgent on the same 10,033-image LFwC set. Each ablation keeps QEMU execution, boot hooks, overlay infrastructure, renderers, and probes runnable while replacing one agent's adaptive decisions with the closest deterministic fallback used by the substrate.

\textbf{State and exposure fidelity.} The ablation results treat persistent state and network exposure as first-order recovery dimensions rather than implementation details. NVRAMAgent admits candidate keys from execution-consulted reads, missing-key traces, defaults and configuration files, static strings, and device/LAN/Web identity evidence. Its typed action interface preserves source and guard fields, while deterministic filters reject malformed fragments, web assets, symbol-like tokens, broad credential-series artifacts, and empty values before overlay materialization. NetworkAgent follows the same evidence discipline by deriving exposure deltas from interface/IP traces, bridge/VLAN/MAC events, service bind addresses, NVRAM LAN evidence, and the \baseline-compatible seed, then rendering only selected interface, address, bridge/VLAN, port, and QEMU-argument tuples.

\textbf{Full-success state/exposure distribution.} We audit the same 5,256 Web-success images used throughout the downstream workflow analysis. NVRAM key files appear on 1,975 images, and nonempty NVRAM materialization covers 1,971 images with 479,440 key entries and 5,320 file-backed entries. The dominant key classes are system/default state (170,925), wireless/region state (119,536), LAN/interface state (105,064), WAN/DNS/gateway state (33,391), service exposure (30,995), identity/model state (13,068), and authentication/access state (11,781). Network exposure artifacts cover 5,191 images, materializing 5,370 exported IP entries, 2,703 QEMU network-argument artifacts, 2,607 bridge/VLAN/interface setup scripts, 161 multi-IP cases, and service-probe records on 2,696 images.

\textbf{Paired full-success effect.} RQ2 measures each role on the 5,256 accepted Web-success executions from the full system. A loss is counted when one of these executions no longer reaches Web under the corresponding deterministic fallback. Replacing PlanAgent loses 2,472 images across 590 brand-device families; replacing NVRAMAgent loses 2,003 images across 462 families, with 1,671 also losing Ping; replacing SearchAgent, NetworkAgent, and FileAgent loses 1,568, 1,558, and 1,525 images, respectively. The NVRAM and Network dependency sets are coupled but distinct. In this split, 1,484 images fail under both removals, 519 only without NVRAMAgent, and 74 only without NetworkAgent. In an ASUS DSL-N10\_C1 case, the full run reconstructs 198 execution-consulted NVRAM keys across LAN, WAN, UPnP, region, and interface state, starts \texttt{/usr/sbin/httpd}, and passes both probes; the no-NVRAM run loses Web reachability. In an ASUS DSL-AC88U case, the full run recovers \texttt{br0} at \texttt{192.168.1.1}, maps the management address through the observed bridge/interface relation, and exposes HTTP to the probe path; the no-Network run loses Web reachability. These cases link aggregate losses to concrete transitions and show corpus-scale state and exposure recovery.

\begin{figure}[t]
  \centering
  \includegraphics[width=0.95\linewidth]{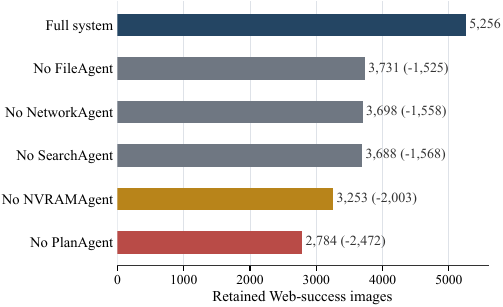}
  \caption{Retained full-system Web successes under leave-one-module ablations; labels on ablated bars show lost full-system successes.}
  \label{fig:rq2_ablation}
\end{figure}

\textbf{Agent contributions.} Figure~\ref{fig:rq2_ablation} reports how many full-system Web successes remain reachable after each leave-one-module replacement. The full system reaches 5,256 images. Replacing PlanAgent with a fixed schedule causes the largest drop, retaining 2,784 of them and losing 2,472. Replacing NVRAMAgent with deterministic defaults retains 3,253 and loses 2,003. Replacing FileAgent, NetworkAgent, and SearchAgent with their deterministic fallbacks retains 3,731, 3,698, and 3,688, respectively. These paired losses show that the agents are complementary. Planning, persistent-state synthesis, network exposure, retrieval, and filesystem/init recovery each contribute measurable gains beyond template-driven fallbacks.

\textbf{Mechanism interpretation.} PlanAgent is the most influential component because environment transitions are order-sensitive. Network exposure before persistent-state recovery, or startup changes without runtime evidence, can create conflicting states. NVRAMAgent confirms the importance of execution-consulted state such as region, device mode, interface names, and authentication defaults, accounting for 2,003 lost Web successes when replaced by deterministic defaults. SearchAgent, NetworkAgent, and FileAgent each account for more than 1,500 lost Web successes, covering long-tail evidence retrieval, log-grounded interface and address exposure, and boot/init recovery. The different loss patterns show that \tool's gain comes from coordinated agent roles validated through execution feedback, rather than from a single permissive rule or isolated artifact edit.

\subsection{RQ3 General-Purpose Coding Agent Baseline}
\label{sec:evaluation:rq3}

\textbf{Baseline protocol.} RQ3 tests whether general coding-agent capability transfers to firmware rehosting without the domain-specific recovery structure used by \tool. We evaluate Claude Code as a single general-purpose coding-agent baseline on the same 10,033-image LFwC set and the same 2,400-second per-image wall-clock budget as \tool. For each image, the baseline receives the same firmware package, extracted workspace, low-level rehosting utilities, run-script context, serial-log evidence, and probe-output format available to the recovery workflow, but it is isolated from \tool's multi-agent orchestrator and typed action interfaces. The comparison uses the same Ping and Web-service predicates defined above, the same guest-IP and HTTP(S) probe logic, and one latest-result record per firmware image for reachability and resource accounting. Under this controlled protocol, Claude Code reaches 729 images by Ping and 545 images by Web, corresponding to 7.27\% and 5.43\%. In contrast, \tool reaches 7,217 images by Ping and 5,256 images by Web.

\begin{table}[t]
  \centering
  \caption{General-Agent Baseline on the 10,033-Image LFwC Corpus}
  \label{tab:rq3_claude}
  {\scriptsize
  \setlength{\tabcolsep}{2.4pt}
  \renewcommand{\arraystretch}{1.08}
  \begin{tabular*}{\linewidth}{@{\extracolsep{\fill}}lccc@{}}
    \toprule
    \textbf{Metric} & \textbf{Claude Code} & \textbf{\tool} & \textbf{Delta / ratio} \\
    \midrule
    Ping reachable & 729 (7.27\%) & 7,217 (71.93\%) & +64.66 pp / 9.90$\times$ \\
    Web reachable & 545 (5.43\%) & 5,256 (52.39\%) & +46.96 pp / 9.64$\times$ \\
    \midrule
    Model tokens/image & 3.61M & 12.4K & 291$\times$ fewer \\
    \midrule
    Shared Web successes & 314 & 314 & 1.00$\times$ \\
    Unique Web successes & 231 & 4,942 & 21.4$\times$ more \\
    Coverage of \tool Web set & 5.97\% & 100\% & 16.8$\times$ \\
    \bottomrule
  \end{tabular*}
  }
\end{table}

\textbf{Success-set overlap.} The Web-success partition in Table~\ref{tab:rq3_claude} further clarifies the difference between generic exploration and domain-specific environment recovery. Only 314 web successes are shared between Claude Code and \tool. Claude Code has 231 unique web successes, but \tool has 4,942 unique web successes. Measured against \tool's web-success set, Claude Code recovers only 5.97\% of the images that \tool can expose as web services. This gap indicates that the task is not simply to generate plausible shell commands or code edits. The system must identify firmware-specific state assumptions, apply bounded artifact transitions through structured interfaces, and accept changes only after execution-grounded validation.

\textbf{Resource profile.} The token profile strengthens this conclusion. Using the same 10,033-image LFwC denominator, Claude Code's latest-result aggregation consumes 36.23B model tokens, corresponding to 3.61M tokens per firmware. \tool's main-run logs consume 12.4 thousand tokens per firmware. Claude Code therefore uses about 291 times more model context while recovering far fewer services.

\textbf{Failure pattern.} The baseline often spends its budget exploring a local workspace without converging on the coupled device environment assumptions that govern firmware startup. In contrast, \tool exposes these assumptions as explicit recovery surfaces covering filesystem/init selection, persistent key--value state, and network exposure. The evidence is cumulative. RQ1 separates \tool from template-driven deterministic rehosting, RQ2 identifies which specialized decisions carry the gain after the shared substrate is held fixed, and RQ3 shows that a strong general coding agent does not recover the same service surface when it lacks the domain action space and feedback loop. Scalable firmware rehosting therefore benefits from a domain-specific control plane that decides when model reasoning may enter the loop, what form its output may take, and how execution results accept or reject that output.

\subsection{RQ4 Support for Downstream Security Workflows}
\label{sec:evaluation:rq4}

\textbf{Workflow scope.} RQ4 evaluates whether web-reachable rehosts preserve enough service behavior for standard downstream security workflows. We analyze all 5,256 LFwC web-success images, covering 622 brand-device families after collapsing firmware-version variants by \texttt{brand} and \texttt{device\_name}. The audit uses service discovery, RouterSploit known-check interaction, and protocol-aware fuzzing, crediting only per-firmware artifacts written by the workflow.

\begin{figure}[!t]
  \centering
  \includegraphics[width=0.88\linewidth]{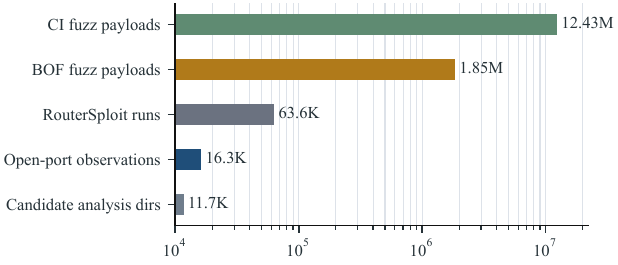}
  \caption{RQ4 Workflow Scale on 5,256 Web-Success Images}
  \label{fig:rq4_full5256}
\end{figure}

\textbf{Audit coverage.} Figure~\ref{fig:rq4_full5256} summarizes the downstream artifact scale. Each audited image contributes service-discovery records, web-fingerprint evidence, RouterSploit transcripts, and protocol-input artifacts. These records bind the recovered endpoint to the tool interaction that exercised it, so RQ4 measures executable workflow evidence rather than raw open-port counts.

\begin{table}[!t]
  \centering
  \caption{LFwC Service Discovery and Web-Fingerprint Coverage}
  \label{tab:rq4_surface}
  {\scriptsize
  \setlength{\tabcolsep}{2.0pt}
  \renewcommand{\arraystretch}{1.06}
  \begin{tabular*}{0.99\linewidth}{@{\extracolsep{\fill}}p{0.42\linewidth}rrr@{}}
    \toprule
    \textbf{Metric} & \textbf{\baseline} & \textbf{Full audit} & \textbf{Gain} \\
    \midrule
    Web-success images analyzed & 2,557 & 5,256 & +2,699 \\
    Nmap-valid images & 2,557 & 5,256 & +2,699 \\
    Web-fingerprint evidence & 2,557 & 5,256 & +2,699 \\
    Detected service records & 8,654 & 16,290 & +7,636 \\
    Distinct service labels & 46 & 57 & +11 \\
    HTTP/HTTPS service images & 2,332 & 4,865 & +2,533 \\
    Telnet service images & 561 & 1,542 & +981 \\
    DNS/domain images & 911 & 1,607 & +696 \\
    UPnP service images & 226 & 358 & +132 \\
    \bottomrule
  \end{tabular*}
  }
\end{table}

\textbf{Service discovery.} Table~\ref{tab:rq4_surface} shows that the recovered environments expose a substantially broader service surface than the deterministic baseline. Across the same LFwC denominator, \tool increases the average number of detected services per firmware from 0.86 to 1.62. The valid service records include protocol labels, endpoints, and available banners or fingerprints, giving downstream tools concrete targets rather than only reachability labels.

\begin{table}[!t]
  \centering
  \caption{RouterSploit 1-Day Validation Results}
  \label{tab:rq4_routersploit}
  {\scriptsize
  \setlength{\tabcolsep}{1.3pt}
  \renewcommand{\arraystretch}{1.04}
  \begin{tabular*}{\linewidth}{@{}l@{\extracolsep{\fill}}rrrl@{}}
    \toprule
    \textbf{Validation layer} & \textbf{Count} & \textbf{Images} & \textbf{Families} & \textbf{Evidence} \\
    \midrule
    RouterSploit audit set & 63,561 & 5,256 & 586 & 33 public 1-day modules \\
    Netgear CVE-2017-5521 & 4,066 & 989 & 819 & 41 positives / 23 images \\
    D-Link CVE-2015-2051 & 1,715 & 646 & 451 & HNAP RCE path \\
    Misfortune Cookie CVE-2014-9222 & 1,138 & 415 & 286 & cookie path \\
    Strict known-check positives & 65 & 35 & 35 & 3 known-check classes \\
    Default-credential positives & 343 & 302 & 275 & FTP/SSH/Telnet auth \\
    All strict positives & 408 & 336 & 309 & module-confirmed outputs \\
    \bottomrule
  \end{tabular*}
  }
\end{table}

\textbf{Known-check interaction.} RouterSploit turns the recovered service surface into executable known-check workflows. We run RouterSploit over the full 5,256-image RQ4 audit set and obtain output and module records for every recovered web-success image. Across this full set, 63,561 module runs span 586 brand-device families and 33 public router modules covering command execution, information disclosure, path traversal, authentication bypass, default credentials, and related checks. Because these modules require service dialogue, request paths, authentication state, or protocol exchanges, positive transcripts validate behavior beyond service discovery.

\textbf{Positive transcripts.} The CVE-linked checks cover Netgear password disclosure (CVE-2017-5521), D-Link HNAP RCE (CVE-2015-2051), and Misfortune Cookie (CVE-2014-9222). The strict subset contains 408 module-confirmed positives across 336 images and 309 families, including 65 known-check positives from exploit-oriented modules and 343 FTP/SSH/Telnet default-credential authentications across 302 images. Manual validation confirms 193 reproducible RouterSploit findings, including 30 known-vulnerability findings and 163 authentication-exposure findings. These results show that the recovered service surface supports stateful validation of known-vulnerability paths and authentication exposures.

\textbf{Protocol-aware testing.} The recovered service surfaces also support protocol-aware input delivery. The fuzzer derives input sites from recovered CGI parameters, forms, HNAP or SOAP fields, UPnP and SSDP identifiers, and cookie-bearing entries. CI mode exercises command-oriented inputs, while BOF mode stresses request and protocol parsers with cyclic or fixed-size buffers. Crash triage uses runtime log tails and deduplicates signals by image, mode, phase, crash kind, marker, and excerpt.

\begin{table}[!t]
  \centering
  \caption{Protocol-Aware Fuzzing Signal Classes}
  \label{tab:rq4_fuzzing}
  {\scriptsize
  \setlength{\tabcolsep}{2.0pt}
  \renewcommand{\arraystretch}{1.05}
  \begin{tabular*}{0.99\linewidth}{@{\extracolsep{\fill}}p{0.34\linewidth}rrp{0.42\linewidth}@{}}
    \toprule
    \textbf{Signal class} & \textbf{Images} & \textbf{Hits} & \textbf{Representative evidence} \\
    \midrule
    CI crash candidates & 1,331 & 4,012 & actions/cookies; SOAP/HNAP; SSDP \\
    BOF crash candidates & 1,522 & 4,591 & 256--12,000B buffers, cyclic markers \\
    Any crash candidate & 1,565 & 8,603 & deduplicated CI/BOF evidence \\
    HTTP/CGI handlers & 554 & 1,263 & httpd, mini\_httpd, uhttpd, CGI \\
    Control/config daemons & 280 & 1,708 & rc, scfgmgr, procd, heartbeat \\
    UPnP/SOAP/SSDP parsers & 64 & 134 & upnp, miniupnpd, SOAP/HNAP fields \\
    Runtime command markers & 31 & 69 & command-dispatch payload markers \\
    \bottomrule
  \end{tabular*}
  }
\end{table}

\textbf{Runtime feedback.} Table~\ref{tab:rq4_fuzzing} reports 4,012 CI crash hits, 4,591 BOF hits, and 1,565 images with at least one image-level crash candidate. The 1,288-image CI/BOF overlap shows that recovered services expose input paths with both semantic reachability and memory-stress feedback, giving downstream testing tools concrete endpoint, protocol, and process context.

\textbf{RQ4 takeaway.} RQ4 shows that \tool turns reachability gains into downstream workflow support. The recovered environments expose broader service surfaces than \baseline, sustain RouterSploit known-check interactions with manually validated findings, and support protocol-aware input delivery with runtime feedback. These results show that evidence-guided environment recovery improves not only Web reachability but also the practical utility of rehosted firmware executions.

\subsection{Threats to Validity and Mitigations}
\label{sec:evaluation:validity}

\textbf{Dataset scope.} LFwC provides metadata for 10,913 firmware records; the executable set contains the 10,033 records whose packages remained downloadable from recorded vendor or archival locations, and all main LFwC comparisons use this denominator.

\textbf{Workflow fidelity.} RQ4 counts service discovery, fuzzing, and RouterSploit evidence only from per-image workflow records, making the evaluation stricter than raw reachability.

\textbf{Baseline scope.} \baseline is the direct deterministic baseline because it supports the same full-image QEMU execution and Ping/Web predicates. Claude Code evaluates the separate question of whether a general-purpose coding agent can substitute for the domain-specific action space under the same denominator, timeout, evidence bundle, and success predicates.

\textbf{LLM variability.} \tool reports outcomes through fixed probes and per-image execution records, while constraining model outputs to typed actions and deterministic artifact renderers. Replications should keep probes and denominators fixed across models.

\section{Related Work}
\label{sec:related}

\textbf{Scalable Firmware Rehosting.} Prior systems improve different parts of the firmware-analysis stack. Firmadyne-style systems construct disk images, infer architecture and kernels, synthesize run scripts, and derive network parameters through deterministic templates~\cite{chen2016towards}. \baseline strengthens this full-image lineage with arbitrated emulation, NVRAM defaults, and network heuristics, making it the closest template-driven deterministic baseline for the Ping/Web reachability protocol used in this work~\cite{kim2020firmae}. FirmGuide and PANDaWAN focus on diagnosing or guiding rehosting failures~\cite{liu2021firmguide,angelakopoulos2024pandawan}, while Greenhouse, FIRMWELL, and user-space dependency-aware rehosting reduce execution complexity by isolating services or user-space dependency environments~\cite{tay2023greenhouse,qin2026firmwell}. \tool targets a different point in this space because it keeps the full-system execution setting but replaces fixed environment templates with evidence-grounded recovery over boot, persistent state, and exposure artifacts.

The distinction is important for scale. Identifying a missing init dependency, state key, or service condition is not enough unless the system can translate that evidence into a repeatable execution environment. \tool makes that path part of the system. Evidence is converted into typed deltas, deltas are applied through deterministic renderers, and the resulting firmware execution is re-probed under the same predicates. This design keeps the evaluation aligned with executable environment recovery rather than isolated failure explanation.

\textbf{Emulation Fidelity.} Hybrid execution combines emulation with hardware for complex peripherals~\cite{zaddach2014avatar,muench2018avatar2}, while modeling and abstraction improve portability and fuzzing effectiveness~\cite{clements2020halucinator,feng2020p2im,cao2020device}. Invalidity-guided inference uses execution failures to discover missing assumptions~\cite{zhou2021automatic}. These works enrich the execution substrate; \tool instead coordinates cross-layer service recovery on scalable whole-system rehosts.

\textbf{Firmware Security Analysis.} Embedded security analyses require reachable services and stable execution for validation and triage. Prior work covers authentication-bypass detection~\cite{shoshitaishvili2015firmalice}, application-driven IoT fuzzing~\cite{chen2018iotfuzzer}, keyword-based bug discovery~\cite{chen2021sharing}, greybox fuzzing~\cite{fioraldi2020aflplusplus}, firmware fuzzing~\cite{scharnowski2022fuzzware,bars2023fuzztruction}, driver/peripheral analysis~\cite{shen2022drifuzz,talebi2018charm}, and tracing/replay~\cite{craig2021pypanda,shoshitaishvili2016sok}. \tool complements these tools by making more firmware executions reachable and analyzable.

\textbf{LLM Agents for Automated Analysis.} Agentic LLM systems support penetration testing~\cite{deng2024pentestgpt}, fuzzing guidance~\cite{meng2024llmguided}, program/binary analysis~\cite{li2023hitchhiker,liu2023llmbinary}, multi-agent coordination~\cite{li2023camel,abramovich2024enigma}, and program-transformation pipelines with retrieval, planning, patch generation, and validation~\cite{seo2025autopatch,pmlr-v267-li25cf,yu2025patchagent}. \tool adapts this paradigm to firmware rehosting by binding agent decisions to typed environment updates, where retrieval supplies evidence, planning orders recovery actions, specialized agents update state and exposure, and reruns with fixed probes validate the resulting artifacts.

Firmware rehosting differs from many code-repair or penetration-testing settings because the target behavior is not specified by a test suite or a live remote service. The desired environment must be reconstructed from partial device assumptions, including init scripts, persistent-state reads, network-interface conventions, and probe responses. \tool therefore uses LLM agents as bounded environment-recovery components rather than open-ended operators. The Claude Code baseline in RQ3 shows why this distinction matters because broad coding ability does not replace domain-specific evidence routing, action schemas, and execution-grounded acceptance.

\section{Discussion}
\label{sec:discussion}

\textbf{Rationale for a Multi-Agent Architecture.} Rehosting failures are cross-layer and sequential, so progress requires coordinated evidence interpretation and constrained actuation. Role specialization separates \emph{what to infer} from \emph{what to change} by assigning Search and planning to symbolic evidence and File, NVRAM, and Network agents to concrete artifact surfaces. The ablations show that removing one role disables a specific recovery channel rather than a monolithic prompt.

\textbf{Security-Workflow Fidelity.} Web reachability is only the entry criterion. RQ4 requires recovered services to sustain enumeration, manual validation of 1-day and credential-path findings, protocol-input delivery, and crash triage, turning reachability into an execution substrate for security analysis.

\textbf{Verifiability.} The unit of evidence is a per-image execution trace rather than a model explanation. Each trace links observations, retrieved support, typed deltas, rendered artifacts, probes, final labels, and downstream tool outputs. This structure makes the reported counts auditable from the same records that drive the recovery loop and keeps failures inspectable at the transition where execution stopped.

\textbf{Limitations and Future Work.} \tool does not fully address peripheral and device-specific I/O modeling, and customized stacks can still fail because of late-stage dependencies such as missing libraries, certificates, or vendor wrappers. Future work can combine environment recovery with hardware-in-the-loop execution~\cite{zaddach2014avatar,muench2018avatar2} and MMIO modeling~\cite{scharnowski2022fuzzware}.

\textbf{Security and Ethical Considerations.} The evaluation runs offline in isolated rehosting environments and does not scan live devices; generated deltas target analysis artifacts, and real-device follow-up should follow responsible disclosure.

\section{Conclusion}
\label{sec:conclusion}

\tool addresses firmware-rehosting brittleness through an evidence-guided multi-agent environment-recovery loop over boot, state, and network layers. Specialized agents interpret runtime evidence and apply typed, replayable artifact deltas. Across LFwC and the public \baseline benchmark, \tool improves service reachability and supports service discovery, manual validation of RouterSploit findings, and protocol-aware fuzzing. The results show that agentic reasoning is effective for firmware rehosting when constrained by evidence, explicit transition interfaces, and repeated execution.

\begingroup
\sloppy
\setlength{\emergencystretch}{3em}
\hfuzz=2pt
\bibliographystyle{IEEEtran}
\bibliography{_reference}

\begin{thebibliography}{10}
\providecommand{\url}[1]{#1}
\csname url@samestyle\endcsname
\providecommand{\newblock}{\relax}
\providecommand{\bibinfo}[2]{#2}
\providecommand{\BIBentrySTDinterwordspacing}{\spaceskip=0pt\relax}
\providecommand{\BIBentryALTinterwordstretchfactor}{4}
\providecommand{\BIBentryALTinterwordspacing}{\spaceskip=\fontdimen2\font plus
\BIBentryALTinterwordstretchfactor\fontdimen3\font minus
  \fontdimen4\font\relax}
\providecommand{\BIBforeignlanguage}[2]{{%
\expandafter\ifx\csname l@#1\endcsname\relax
\typeout{** WARNING: IEEEtran.bst: No hyphenation pattern has been}%
\typeout{** loaded for the language `#1'. Using the pattern for}%
\typeout{** the default language instead.}%
\else
\language=\csname l@#1\endcsname
\fi
#2}}
\providecommand{\BIBdecl}{\relax}
\BIBdecl

\bibitem{chen2016towards}
D.~D. Chen, M.~Woo, D.~Brumley, and M.~Egele, ``Towards automated dynamic
  analysis for linux-based embedded firmware.'' in \emph{NDSS}, 2016.

\bibitem{kim2020firmae}
M.~Kim, D.~Kim, E.~Kim, S.~Kim, Y.~Jang, and Y.~Kim, ``Firmae: Towards
  large-scale emulation of iot firmware for dynamic analysis,'' in
  \emph{Proceedings of the 36th Annual Computer Security Applications
  Conference}, 2020, pp. 733--745.

\bibitem{liu2021firmguide}
Q.~Liu, C.~Zhang, L.~Ma, M.~Jiang, Y.~Zhou, L.~Wu, W.~Shen, X.~Luo, Y.~Liu, and
  K.~Ren, ``Firmguide: Boosting the capability of rehosting embedded linux
  kernels through model-guided kernel execution,'' in \emph{2021 36th IEEE/ACM
  International Conference on Automated Software Engineering (ASE)}, 2021, pp.
  792--804.

\bibitem{angelakopoulos2024pandawan}
I.~Angelakopoulos, G.~Stringhini, and M.~Egele, ``Pandawan: quantifying
  progress in linux-based firmware rehosting,'' in \emph{33rd USENIX Security
  Symposium (USENIX Security 24)}, 2024, pp. 5859--5876.

\bibitem{johnson2021jetset}
E.~Johnson, M.~Bland, Y.~Zhu, J.~Mason, S.~Checkoway, S.~Savage, and
  K.~Levchenko, ``Jetset: Targeted firmware rehosting for embedded systems,''
  in \emph{30th USENIX Security Symposium}, 2021, pp. 321--338.

\bibitem{tay2023greenhouse}
H.~J. Tay, K.~Zeng, J.~M. Vadayath, A.~S. Raj, A.~Dutcher, T.~Reddy, W.~Gibbs,
  Z.~L. Basque, F.~Dong, Z.~Smith \emph{et~al.},
  ``Greenhouse:$\{$Single-Service$\}$ rehosting of $\{$Linux-Based$\}$ firmware
  binaries in $\{$User-Space$\}$ emulation,'' in \emph{32nd USENIX Security
  Symposium}, 2023, pp. 5791--5808.

\bibitem{qin2026firmwell}
C.~Qin, C.~Zhang, Y.~Zheng, P.~Liu, J.~Zhang, Y.~Li, W.~Zhang, Y.~Liu, and
  L.~Sun, ``User-space dependency-aware rehosting for linux-based firmware
  binaries,'' in \emph{Proceedings of the Network and Distributed System
  Security Symposium}, 2026.

\bibitem{muench2018avatar2}
M.~Muench, D.~Nisi, A.~Francillon, and D.~Balzarotti, ``{Avatar}$^2$: A
  multi-target orchestration platform,'' in \emph{Proceedings 2018 Workshop on
  Binary Analysis Research}, 2018.

\bibitem{clements2020halucinator}
A.~A. Clements, E.~Gustafson, T.~Scharnowski, P.~Grosen, D.~Fritz, C.~Kruegel,
  G.~Vigna, S.~Bagchi, and M.~Payer, ``$\{$HALucinator$\}$: Firmware re-hosting
  through abstraction layer emulation,'' in \emph{29th USENIX Security
  Symposium (USENIX Security 20)}, 2020, pp. 1201--1218.

\bibitem{feng2020p2im}
B.~Feng, A.~Mera, and L.~Lu, ``$\{$P2IM$\}$: Scalable and hardware-independent
  firmware testing via automatic peripheral interface modeling,'' in \emph{29th
  USENIX Security Symposium (USENIX Security 20)}, 2020, pp. 1237--1254.

\bibitem{scharnowski2022fuzzware}
T.~Scharnowski, N.~Bars, M.~Schloegel, E.~Gustafson, M.~Muench, G.~Vigna,
  C.~Kruegel, T.~Holz, and A.~Abbasi, ``Fuzzware: Using precise $\{$MMIO$\}$
  modeling for effective firmware fuzzing,'' in \emph{31st USENIX Security
  Symposium}, 2022, pp. 1239--1256.

\bibitem{bellard2005qemu}
F.~Bellard, ``{QEMU}, a fast and portable dynamic translator,'' in \emph{USENIX
  Annual Technical Conference, FREENIX Track}, 2005, pp. 41--46.

\bibitem{fasano2021sok}
A.~Fasano, T.~Ballo, M.~Muench, T.~Leek, A.~Bulekov, B.~Dolan-Gavitt, M.~Egele,
  A.~Francillon, L.~Lu, N.~Gregory \emph{et~al.}, ``Sok: Enabling security
  analyses of embedded systems via rehosting,'' in \emph{Proceedings of the
  2021 ACM Asia Conference on Computer and Communications Security}, 2021, pp.
  687--701.

\bibitem{gustafson2019toward}
E.~Gustafson, M.~Muench, C.~Spensky, N.~Redini, A.~Machiry, Y.~Fratantonio,
  D.~Balzarotti, A.~Francillon, Y.~R. Choe, C.~Kruegel \emph{et~al.}, ``Toward
  the analysis of embedded firmware through automated re-hosting,'' in
  \emph{22nd International Symposium on Research in Attacks, Intrusions and
  Defenses (RAID 2019)}, 2019, pp. 135--150.

\bibitem{redini2020karonte}
N.~Redini, A.~Machiry, R.~Wang, C.~Spensky, A.~Continella, Y.~Shoshitaishvili,
  C.~Kruegel, and G.~Vigna, ``Karonte: Detecting insecure multi-binary
  interactions in embedded firmware,'' in \emph{2020 IEEE Symposium on Security
  and Privacy (SP)}, 2020, pp. 1544--1561.

\bibitem{zaddach2014avatar}
J.~Zaddach, L.~Bruno, A.~Francillon, D.~Balzarotti \emph{et~al.}, ``Avatar: A
  framework to support dynamic security analysis of embedded systems'
  firmwares.'' in \emph{NDSS}, 2014, pp. 1--16.

\bibitem{lewis2020rag}
P.~Lewis, E.~Perez, A.~Piktus, F.~Petroni, V.~Karpukhin, N.~Goyal,
  H.~K{\"u}ttler, M.~Lewis, W.-t. Yih, T.~Rockt{\"a}schel \emph{et~al.},
  ``Retrieval-augmented generation for knowledge-intensive nlp tasks,''
  \emph{Advances in neural information processing systems}, vol.~33, pp.
  9459--9474, 2020.

\bibitem{wei2022cot}
J.~Wei, X.~Wang, D.~Schuurmans, M.~Bosma, F.~Xia, E.~Chi, Q.~V. Le, D.~Zhou
  \emph{et~al.}, ``Chain-of-thought prompting elicits reasoning in large
  language models,'' \emph{Advances in neural information processing systems},
  vol.~35, pp. 24\,824--24\,837, 2022.

\bibitem{shinn2024reflexion}
N.~Shinn, F.~Cassano, A.~Gopinath, K.~Narasimhan, and S.~Yao, ``Reflexion:
  Language agents with verbal reinforcement learning,'' \emph{Advances in
  Neural Information Processing Systems}, vol.~36, pp. 8634--8652, 2023.

\bibitem{li2023camel}
G.~Li, H.~Hammoud, H.~Itani, D.~Khizbullin, and B.~Ghanem, ``{CAMEL}:
  Communicative agents for ``mind'' exploration of large language model
  society,'' \emph{Advances in Neural Information Processing Systems}, vol.~36,
  pp. 51\,991--52\,008, 2023.

\bibitem{deng2024pentestgpt}
G.~Deng, Y.~Liu, V.~Mayoral-Vilches, P.~Liu, Y.~Li, Y.~Xu, T.~Zhang, Y.~Liu,
  M.~Pinzger, and S.~Rass, ``$\{$PentestGPT$\}$: Evaluating and harnessing
  large language models for automated penetration testing,'' in \emph{33rd
  USENIX Security Symposium}, 2024, pp. 847--864.

\bibitem{meng2024llmguided}
R.~Meng, M.~Mirchev, M.~B{\"o}hme, and A.~Roychoudhury, ``Large language model
  guided protocol fuzzing,'' in \emph{Proceedings 2024 Network and Distributed
  System Security Symposium}, 2024.

\bibitem{li2023hitchhiker}
H.~Li, Y.~Hao, Y.~Zhai, and Z.~Qian, ``The hitchhiker's guide to program
  analysis: A journey with large language models,'' \emph{arXiv preprint
  arXiv:2308.00245}, 2023.

\bibitem{liu2023llmbinary}
P.~Liu, C.~Sun, Y.~Zheng, X.~Feng, C.~Qin, Y.~Wang, Z.~Li, and L.~Sun,
  ``Harnessing the power of llm to support binary taint analysis,'' \emph{arXiv
  preprint arXiv:2310.08275}, 2023.

\bibitem{helmke2024lfwc}
R.~Helmke, E.~Padilla, and N.~Aschenbruck, ``Mens sana in corpore sano: Sound
  firmware corpora for vulnerability research,'' in \emph{Proceedings of the
  2025 Network and Distributed System Security Symposium (NDSS)}, 2025.

\bibitem{cao2020device}
C.~Cao, L.~Guan, J.~Ming, and P.~Liu, ``Device-agnostic firmware execution is
  possible: A concolic execution approach for peripheral emulation,'' in
  \emph{Proceedings of the 36th Annual Computer Security Applications
  Conference}, 2020, pp. 746--759.

\bibitem{zhou2021automatic}
W.~Zhou, L.~Guan, P.~Liu, and Y.~Zhang, ``Automatic firmware emulation through
  invalidity-guided knowledge inference,'' in \emph{30th USENIX Security
  Symposium (USENIX Security 21)}, 2021, pp. 2007--2024.

\bibitem{shoshitaishvili2015firmalice}
Y.~Shoshitaishvili, R.~Wang, C.~Hauser, C.~Kruegel, and G.~Vigna,
  ``Firmalice-automatic detection of authentication bypass vulnerabilities in
  binary firmware.'' in \emph{NDSS}, 2015.

\bibitem{chen2018iotfuzzer}
J.~Chen, W.~Diao, Q.~Zhao, C.~Zuo, Z.~Lin, X.~Wang, W.~C. Lau, M.~Sun, R.~Yang,
  and K.~Zhang, ``Iotfuzzer: Discovering memory corruptions in iot through
  app-based fuzzing.'' in \emph{NDSS}, 2018, pp. 1--15.

\bibitem{chen2021sharing}
L.~Chen, Y.~Wang, Q.~Cai, Y.~Zhan, H.~Hu, J.~Linghu, Q.~Hou, C.~Zhang, H.~Duan,
  and Z.~Xue, ``Sharing more and checking less: Leveraging common input
  keywords to detect bugs in embedded systems,'' in \emph{30th USENIX Security
  Symposium (USENIX Security 21)}, 2021, pp. 303--319.

\bibitem{fioraldi2020aflplusplus}
A.~Fioraldi, D.~Maier, H.~Ei{\ss}feldt, and M.~Heuse, ``{AFL++}: Combining
  incremental steps of fuzzing research,'' in \emph{14th USENIX Workshop on
  Offensive Technologies (WOOT 20)}, 2020.

\bibitem{bars2023fuzztruction}
N.~Bars, M.~Schloegel, T.~Scharnowski, N.~Schiller, and T.~Holz,
  ``Fuzztruction: Using fault injection-based fuzzing to leverage implicit
  domain knowledge,'' in \emph{32nd USENIX Security Symposium (USENIX Security
  23)}, 2023, pp. 1847--1864.

\bibitem{shen2022drifuzz}
Z.~Shen, R.~Roongta, and B.~Dolan-Gavitt, ``Drifuzz: Harvesting bugs in device
  drivers from golden seeds,'' in \emph{31st USENIX Security Symposium (USENIX
  Security 22)}, 2022, pp. 1275--1290.

\bibitem{talebi2018charm}
S.~M.~S. Talebi, H.~Tavakoli, H.~Zhang, Z.~Zhang, A.~A. Sani, and Z.~Qian,
  ``Charm: Facilitating dynamic analysis of device \allowbreak drivers of
  mobile systems,'' in \emph{27th USENIX Security Symposium (USENIX Security
  18)}, 2018, pp. 291--307.

\bibitem{craig2021pypanda}
L.~Craig, A.~Fasano, T.~Ballo, T.~Leek, B.~Dolan-Gavitt, and W.~Robertson,
  ``{PyPANDA}: Taming the {PANDA}monium of whole system dynamic analysis,'' in
  \emph{Proceedings 2021 Workshop on Binary Analysis Research}, 2021.

\bibitem{shoshitaishvili2016sok}
Y.~Shoshitaishvili, R.~Wang, C.~Salls, N.~Stephens, M.~Polino, A.~Dutcher,
  J.~Grosen, S.~Feng, C.~Hauser, C.~Kruegel \emph{et~al.}, ``Sok:(state of) the
  art of war: Offensive techniques in binary analysis,'' in \emph{2016 IEEE
  Symposium on Security and Privacy (SP)}, 2016, pp. 138--157.

\bibitem{abramovich2024enigma}
T.~Abramovich, M.~Udeshi, M.~Shao, K.~Lieret, H.~Xi, K.~Milner, S.~Jancheska,
  J.~Yang, C.~E. Jimenez, F.~Khorrami \emph{et~al.}, ``{EnIGMA}: Enhanced
  interactive generative model agent for ctf challenges,'' \emph{arXiv preprint
  arXiv:2409.16165}, 2024.

\bibitem{seo2025autopatch}
M.~Seo, W.~Choi, M.~You, and S.~Shin, ``Autopatch: Multi-agent framework for
  patching real-world cve vulnerabilities,'' \emph{arXiv preprint
  arXiv:2505.04195}, 2025.

\bibitem{pmlr-v267-li25cf}
H.~Li, Y.~Tang, S.~Wang, and W.~Guo, ``Patchpilot: A cost-efficient software
  engineering agent with early attempts on formal verification,'' \emph{arXiv
  preprint arXiv:2502.02747}, 2025.

\bibitem{yu2025patchagent}
Z.~Yu, Z.~Guo, Y.~Wu, J.~Yu, M.~Xu, D.~Mu, Y.~Chen, and X.~Xing,
  ``{PATCHAGENT}: A practical program repair agent mimicking human expertise,''
  in \emph{34th USENIX Security Symposium (USENIX Security 25)}, 2025, pp.
  4381--4400.

\end{thebibliography}
\endgroup

\end{document}